\documentclass[reprint,superscriptaddress,aps,prl]{revtex4-1}
\usepackage{amsmath,epsfig,amssymb,subfigure,bm,dsfont}
\usepackage{lipsum} 
\usepackage{color}
\usepackage{graphicx}
\usepackage{dcolumn}
\usepackage{bm}
\usepackage{bbold}
\usepackage{amsfonts,color}
\usepackage{amsmath}
\newcommand{\sslash}{\mathbin{/\mkern-6mu/}}

\newcounter{lastnote}

\begin{document}
\title{Long-distance propagation of high-velocity antiferromagnetic spin waves}
\date{\today}
\author{Hanchen Wang}
\thanks{These authors contributed equally to this work.}
\affiliation{%
	Fert Beijing Institute, MIIT Key Laboratory of Spintronics, School of Integrated Circuit Science and Engineering, Beihang University, Beijing 100191, China
}
\affiliation{%
	International Quantum Academy, Shenzhen 518048, China
}%
\affiliation{%
	Department of Materials, ETH Zurich, Zurich 8093, Switzerland
}%
\author{Rundong Yuan}
\thanks{These authors contributed equally to this work.}
\affiliation{%
	Fert Beijing Institute, MIIT Key Laboratory of Spintronics, School of Integrated Circuit Science and Engineering, Beihang University, Beijing 100191, China
}
\author{Yongjian Zhou}
\thanks{These authors contributed equally to this work.}
\affiliation{%
	Key Laboratory of Advanced Materials (MOE), School of Materials Science and Engineering, Tsinghua University, Beijing 100084, China
}
\author{Yuelin Zhang}
\affiliation{%
	Fert Beijing Institute, MIIT Key Laboratory of Spintronics, School of Integrated Circuit Science and Engineering, Beihang University, Beijing 100191, China
}%
\author{Jilei Chen}
\affiliation{Shenzhen Institute for Quantum Science and Engineering, Southern University of Science and Technology, Shenzhen 518055, China}
\affiliation{%
	International Quantum Academy, Shenzhen 518048, China
}%

\author{Song~Liu}
\affiliation{%
	International Quantum Academy, Shenzhen 518048, China
}%
\affiliation{Shenzhen Institute for Quantum Science and Engineering, Southern University of Science and Technology, Shenzhen 518055, China}

\author{Hao Jia}
\affiliation{%
	International Quantum Academy, Shenzhen 518048, China
}%
\affiliation{Shenzhen Institute for Quantum Science and Engineering, Southern University of Science and Technology, Shenzhen 518055, China}
\author{Dapeng Yu}
\affiliation{%
	International Quantum Academy, Shenzhen 518048, China
}%
\affiliation{Shenzhen Institute for Quantum Science and Engineering, Southern University of Science and Technology, Shenzhen 518055, China}
\author{Jean-Philippe Ansermet}
\email{jean-philippe.ansermet@epfl.ch}
\affiliation{%
Institute of Physics, Ecole Polytechnique F\'ed\'erale de Lausanne (EPFL), 1015, Lausanne, Switzerland
}
\affiliation{Shenzhen Institute for Quantum Science and Engineering, Southern University of Science and Technology, Shenzhen 518055, China}
\author{Cheng Song}
\email{songcheng@mail.tsinghua.edu.cn}
\affiliation{%
Key Laboratory of Advanced Materials (MOE), School of Materials Science and Engineering, Tsinghua University, Beijing 100084, China
}
\author{Haiming Yu}
\email{haiming.yu@buaa.edu.cn}
\affiliation{%
Fert Beijing Institute, MIIT Key Laboratory of Spintronics, School of Integrated Circuit Science and Engineering, Beihang University, Beijing 100191, China
}
\affiliation{%
International Quantum Academy, Shenzhen 518048, China
}%

\begin{abstract}
We report on coherent propagation of antiferromagnetic (AFM) spin waves over a long distance ($\sim$10 $\mu$m) at room temperature in a canted AFM $\alpha$-Fe$_2$O$_3$ with the Dzyaloshinskii-Moriya interaction (DMI). Unprecedented high group velocities (up to 22.5 km/s) are characterized by microwave transmission using all-electrical spin wave spectroscopy. We derive analytically AFM spin-wave dispersion in the presence of the DMI which accounts for our experimental results. The AFM spin waves excited by nanometric coplanar waveguides with large wavevectors enter the exchange regime and follow a quasi-linear dispersion relation. Fitting of experimental data with our theoretical model yields an AFM exchange stiffness length of 1.7~$\rm \AA$. Our results provide key insights on AFM spin dynamics and demonstrate high-speed functionality for AFM magnonics.
	
\end{abstract}

\maketitle

\begin{figure}
	\includegraphics[width=86mm]{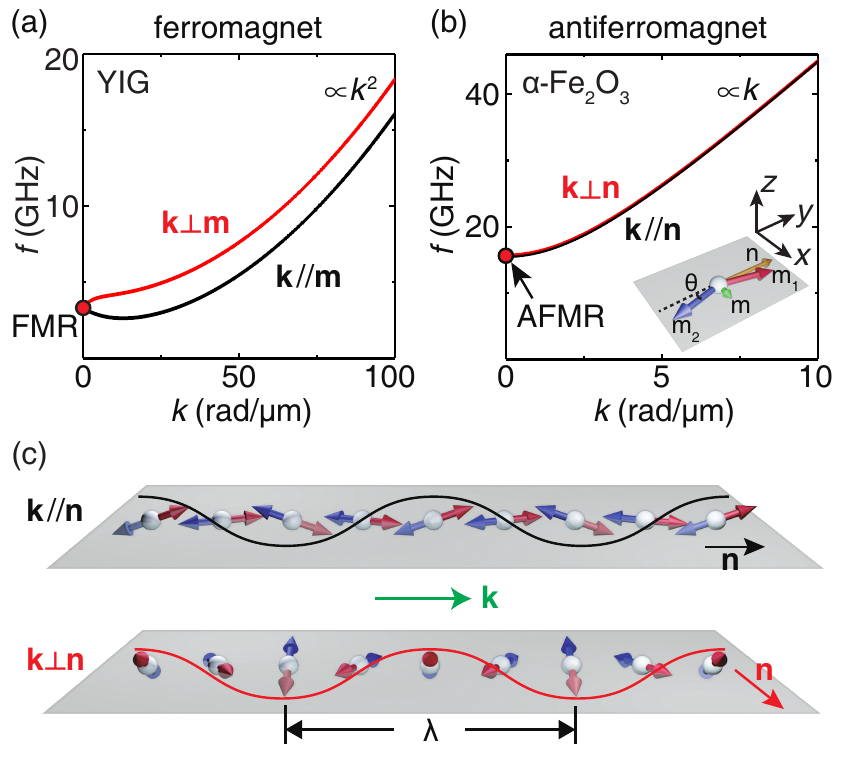}
	\caption{(a) Ferromagnetic-type spin-wave dispersion for a 200 nm-thick YIG film. The $\mathbf{k\perp m}$ and $\mathbf{k\sslash m}$ modes are separated due to the dipolar interaction. The $k=0$ mode is the ferromagnetic resonance (FMR). Exchange-dominated spin waves follow a quadratic $k^2$ relation. (b) Spin-wave dispersion of a canted antiferromagnet $\alpha$-Fe$_2$O$_3$. The $\mathbf{k\perp n}$  and $\mathbf{k\sslash n}$ modes as illustrated in (c) are degenerate and adhere to a linear $k$ dependence in exchange regime. The black arrow marks the antiferromagnetic resonance (AFMR) with $k=0$. Inset illustrates the canting of two sublattices induced by the DMI. (c) Illustrations of exchange spin waves in an easy-plane antiferromagnet at two different configurations, namely, wavevector $\mathbf{k}$ parallel and perpendicular to N\'eel vector $\mathbf{n}$.}
	\label{fig1}
\end{figure}

Spin waves or magnons~\cite{Dirk2010,Chu2015,Pirro2021,Barman2021} as collective spin excitations can transport coherent spin information in a magnetic media over long distances~\cite{vanWees2015,Liu2018,Lebrun2018} without suffering from Ohmic loss, and are therefore promising for magnon-based computing with low-power consumption~\cite{Csaba2016,Chu2022}. So far, an overwhelming majority of magnonic research are conducted in ferro-(ferri-)magnetic material systems, such as yttrium iron garnet (YIG)~\cite{Serga2010,Chang2014,Qin2018,QWang2019,Yu2014}, permalloy~\cite{Vla2008,Neu2010,Ade2016,Wagner2016} and magnetic multilayers~\cite{JHan2019,Ono2020,JLi2022}. In ferromagnetic (FM) materials, long wavelength spin waves are predominantly affected by dipolar interactions, resulting in Damon-Eshbach (DE) and Backward-volume (BV) modes with distinct configurations of the magnetization ($\mathbf{m}$) and wavevector ($\mathbf{k}$) and non-degenerate dispersions [Fig.~\ref{fig1}(a)]. This anisotropy hinders spin waves from propagating through a curved circuit~\cite{Vogt2012} and leads to vulnerability to external field disturbances. Thus, it is desirable to excite high-$k$ exchange spin waves in ferromagnets with short wavelengths that are substantially less anisotropic. Exchange spin waves in ferromagnets~\cite{Slavin1986} follows a parabolic dispersion relation suggesting an increasing group velocity for higher $k$. It is extremely challenging to excite exchange spin waves with a record high velocity around 1 km/s and a wavelength below 100 nm~\cite{Liu2018,Sluka2019,Che2020}.

In antiferromagnetic (AFM) materials, these challenges and obstacles are inherently neutralized because spin waves are insensitive to disturbing magnetic fields~\cite{Baltz2018} and can propagate with higher velocities~\cite{Hort2021}. However, new challenges arise as antiferromagnets have zero net moment~\cite{Jung2016}. In addition, antiferromagnetic spin waves fall typically in the THz or sub-THz frequency regime~\cite{Gom2018,qiu2022manipulating} and are so far mostly excited with an optical method~\cite{Kamp2011,Hort2021}, that is difficult to integrate with on-chip magnonic devices. All-electrical excitation and detection of coherent AFM spin waves~\cite{Hicken2020} are highly desired for magnonics, but remain challenging. 
Recently, advanced microwave technology based on solid-state extenders enabled frequency multiplication of conventional GHz source into sub-THz generators for all-electrical AFM magnon excitation~\cite{Casp2016,JShi2020}. Until now, coherent AFM spin waves are electrically excited only with $k=0$, i.e., antiferromagnetic resonance (AFMR)~\cite{Kamp2011,Casp2016,JShi2020,boventer2022antiferromagnetic} (e.g. black arrow in Fig.~\ref{fig1}(b)), which has zero group velocity in a canted AFM~\cite{Rezende2019}. High-velocity propagating AFM exchange spin waves with electrical excitation has not been realized so far.

In this Letter, we experimentally demonstrate coherent propagation of AFM exchange spin waves over a long distance (10 $\mu$m) in $\alpha$-Fe$_2$O$_3$ with a high group velocity (22.5 km/s) at room temperature. The velocity is approximately one order of magnitude higher than that of FM exchange spin waves~\cite{Liu2018,Sluka2019,Che2020}. $\alpha$-Fe$_2$O$_3$ also known as hematite is an insulating antiferromagnet~\cite{Morin1951,Besser1967,Jani2021,chmiel2018,wittmann2022role,dos2020modeling,Liu2022} with ultra-low magnetic damping ($\sim 10^{-5}$)~\cite{Lebrun2020} and high N\'eel temperature ($\sim$960 K)~\cite{Marcin2022}. At room temperature (above its Morin temperature $T_{\text{M}}\simeq260$ K), $\alpha$-Fe$_2$O$_3$ is in an easy-plane antiferromagnetic phase, where its N\'eel vector $\mathbf{n}$ lies in the plane [Fig.~\ref{fig1}(c)] normal to the corundum crystal $c$ axis~\cite{HLWang2021}. The bulk Dzyaloshinskii-Moriya interaction (DMI) in $\alpha$-Fe$_2$O$_3$ induces a small canted moment~\cite{Boventer2021} as shown in the inset of Fig.~\ref{fig1}(b). The weak canted moment ($\sim$1.2 emu/cm$^3$) existing in the easy plane is measured by vibrating sample magnetometer as shown in the Section~I in Supplementary Material (SM)~\cite{SI}. Although the canted moment is negligibly weak as a net magnetic moment ($\sim$1$\%$ of YIG magnetic moment), it facilitates AFMR excitation with conventional microwave antennas~\cite{HLWang2021,Boventer2021}. As the easy-plane anisotropy is remarkably small ($H_{\text{a}} \sim$0.06~mT), the AFMR frequency drops to around 20~GHz, which is accessible using conventional microwave techniques. The negligible easy-plane anisotropy also allows the N\'eel $\mathbf{n}$ to rotate freely in plane with respect to the spin-wave wavevector $\mathbf{k}$, as illustrated in Fig.~\ref{fig1}(c) for $\mathbf{k} \sslash \mathbf{n}$ and $\mathbf{k}\perp\mathbf{n}$ as examples. Unlike in ferromagnets where $\mathbf{k}\perp\mathbf{m}$ (DE) and $\mathbf{k}\sslash\mathbf{m}$ (BV) behave differently [Fig.~\ref{fig1}(b)], spin waves in antiferromagents are degenerate for $\mathbf{k}\perp\mathbf{n}$ and $\mathbf{k}\sslash\mathbf{n}$ [Fig.~\ref{fig1}(c)] and any intermediate angle because the AFM spin-wave dispersion is not affected by dipolar interaction but fully determined by exchange interaction. Recently, Boventer $et$ $al.$~\cite{Boventer2021} have theoretically derived the AFMR ($k=0$) formula for $\alpha$-Fe$_2$O$_3$. However, literature on the spin-wave dispersion for an easy-plane antiferromagnet with DMI remains elusive. 

\begin{figure*}
\centering
	\includegraphics[width=172mm]{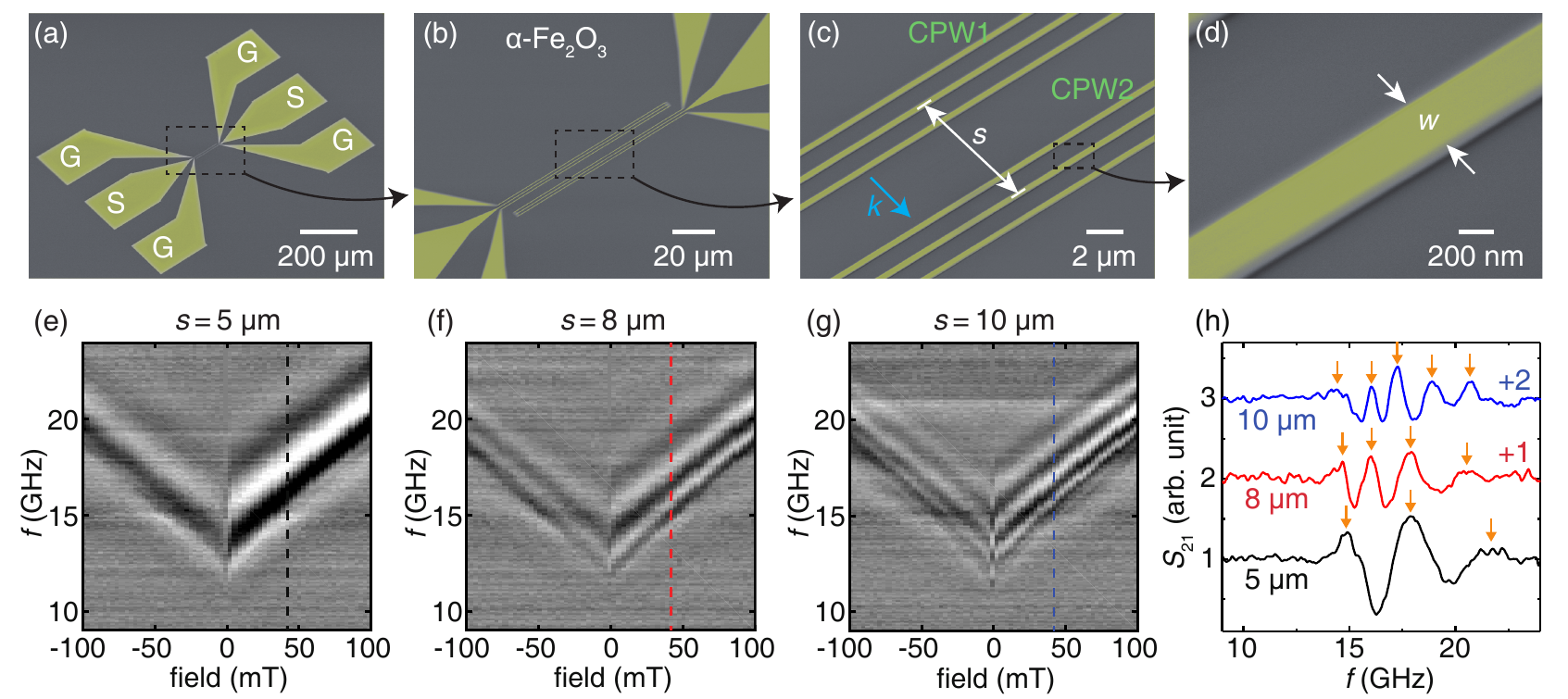}
	\caption{(a) A global scanning electron microscopic (SEM) image of two coplanar waveguide (CPW) antennas integrated on $\alpha$-Fe$_2$O$_3$ with a ground-signal-ground (GSG) design. The pitch of the contact pad is 250$\,\mu$m, compatible with microwave probes. The gray background represents $\alpha$-Fe$_2$O$_3$ substrate. (b) SEM image within the black dashed square area in (a). (c) SEM-resolved image of a CPW. The center-to-center distance $s$ between CPW1 and CPW2 for this device is 8$\,\mu$m. The blue arrow indicate the spin-wave wavevector $k$. The yellow-rendered parts are the gold conducting lines whose width $w\,=\,380\,$nm is characterized by the further close-up image in (d). (e)-(g) Spin-wave transmission spectra $S_{12}$ measured by a vector network analyzer (imaginary part of the $S$ parameter) and plotted as a function of applied magnetic field for three devices with different propagation distances $s=5,\,8,\,10\,\mu$m. (h) Lineplots extracted at 45$\,$mT for propagation distances $s$. Spectra are shifted for clarity. Orange arrows highlight the peak positions.
	}
	\label{fig2}
\end{figure*}
Let us first derive and discuss the spin-wave dispersion for an antiferromagnet with DMI-induced canting and easy-plane anisotropy like that of $\alpha$-Fe$_2$O$_3$ [Fig.~\ref{fig1}(b)]. We consider a one-dimensional spin chain with two sublattices $\textbf{m}_1$ and $\textbf{m}_2$ that are antiferromagnetically coupled and confined in the easy plane. In the AFM system, the exchange energy, Zeeman energy, anisotropy energy and Dzyaloshinskii-Moriya (DM) energy constitute the total free energy, from which we obtain the equations of motion describing the dynamics of two spin sublattices in a mean-field approximation (see Sec.~II in SM~\cite{SI}),
\begin{widetext}
	\begin{equation}\label{EOM}
		\left\{
		\begin{aligned}
			\frac{d\textbf{m}_{1}}{dt}=-\gamma\mu_{0}\textbf{m}_{1}\times\left[\textbf{H}_{0}-H_{\text{ex}}\textbf{m}_{2}-\frac{1}{2}H_{\rm ex}a^2_{\text{ex}}\nabla^{2}\textbf{m}_{2}-H_{\text{A}}\left(\textbf{m}_{1}\cdot\hat{\textbf{z}}\right)\hat{\textbf{z}}-H_{\text{a}}\left(\textbf{m}_{1}\cdot\hat{\textbf{y}}\right)\hat{\textbf{y}}+H_{\mathrm{DM}}\left(\textbf{m}_{2}\times\hat{\textbf{z}}\right)\right]
			\\
			\frac{d\textbf{m}_{2}}{dt}=-\gamma\mu_{0}\textbf{m}_{2}\times\left[\textbf{H}_{0}-H_{\text{ex}}\textbf{m}_{1}-\frac{1}{2}H_{\rm ex}a^2_{\text{ex}}\nabla^{2}\textbf{m}_{1}-H_{\text{A}}\left(\textbf{m}_{1}\cdot\hat{\textbf{z}}\right)\hat{\textbf{z}}-H_{\text{a}}\left(\textbf{m}_{2}\cdot\hat{\textbf{y}}\right)\hat{\textbf{y}}-H_{\mathrm{DM}}\left(\textbf{m}_{1}\times\hat{\textbf{z}}\right)\right]
		\end{aligned}
	    \right.,
	\end{equation}
\end{widetext}
where $\textbf{H}_{0}$ is the external magnetic field, $H_{\text{ex}}$ is the strength of the exchange field, $a_{\text{ex}}$ is the exchange stiffness length (see Sec.~II in SM~\cite{SI}), $H_{\text{A}}$ ($H_{\text{a}}$) is the out-of-plane (in-plane) anisotropy and $H_{\text{DM}}$ is the DM effective field. The exchange stiffness term consisting of the $H_{\text{ex}}$ and $a_{\text{ex}}$ is discussed in the SM Table I~\cite{SI} with a comparison between ferro-(ferri-)magnetic~\cite{Turner1960} and antiferromagnetic models~\cite{Rezende2016}. Coordinate axes are defined in the inset of Fig.~\ref{fig1}(b). Considering a small canting angle $\theta$ induced by the DMI, the cartesian coordinate defined with $\textbf{n}$ and $\textbf{m}$ is subsequently transformed to align with $\textbf{m}_1$ or $\textbf{m}_2$ in order to deduce the dynamics of the sublattices. By extracting the eigenfrequencies of Eq.~\ref{EOM}, one derives the dispersion relations for both acoustic (low-frequency) and optical (high-frequency) AFM magnon modes (see Sec.~II in SM~\cite{SI}). Since only the low-frequency one is relevant for our experiments, we present its spin-wave dispersion here as
\begin{equation}\label{AFM_dispersion}  
	f=\frac{\gamma\mu_0}{2\pi}\sqrt{H_0(H_0+H_{\text{DM}})+2H_{\text{a}}H_{\text{ex}}+H^2_{\text{ex}}a^2_{\rm ex}k^2},
\end{equation} 
where $\gamma$ is the gyromagnetic ratio and $\mu_{0}$ is the vacuum permeability. If on the one dd we take $k=0$, the last term underneath the square root vanishes and thereby it reduces to the uniform AFMR of a canted antiferromagnet as studied by Boventer~$et~al.$~\cite{Boventer2021}. On the other hand, if we do not introduce the DMI in the system, the first term beneath the square root disappears and hence Eq.~\ref{AFM_dispersion} becomes essentially the same as the dispersion for an easy-plane AFM such as NiO as analyzed by Rezende~$et~al.$~\cite{Rezende2016} in the absence of DMI. In Fig.~\ref{fig1}(b), we plot the spin-wave dispersion for the low-frequency mode in $\alpha$-Fe$_2$O$_3$ based on Eq.~\ref{fig2} with $\mu_0 H_{0}=60$~mT. Here we take the exchange field $\mu_0 H_{\text{ex}}=1040$~T from Ref.~\cite{Besser1967}. By fitting the field-dependent AFMR in a flip-chip measurement ($k=0$, see Sec.~III in SM~\cite{SI}), we extract the effective DM field $\mu_0H_{\rm DM}=2.7$~T and $\mu_0 H_{\rm a}\,=\,0.067$~mT. These values are close to those in Refs.~\cite{Boventer2021,HLWang2021} and are adapted in plotting the dispersion in Fig.~\ref{fig1}(b). Thus, the calculated dispersion relation predicts a group velocity of up to 26.5~km/s.

\begin{figure}
	\includegraphics[width=86mm]{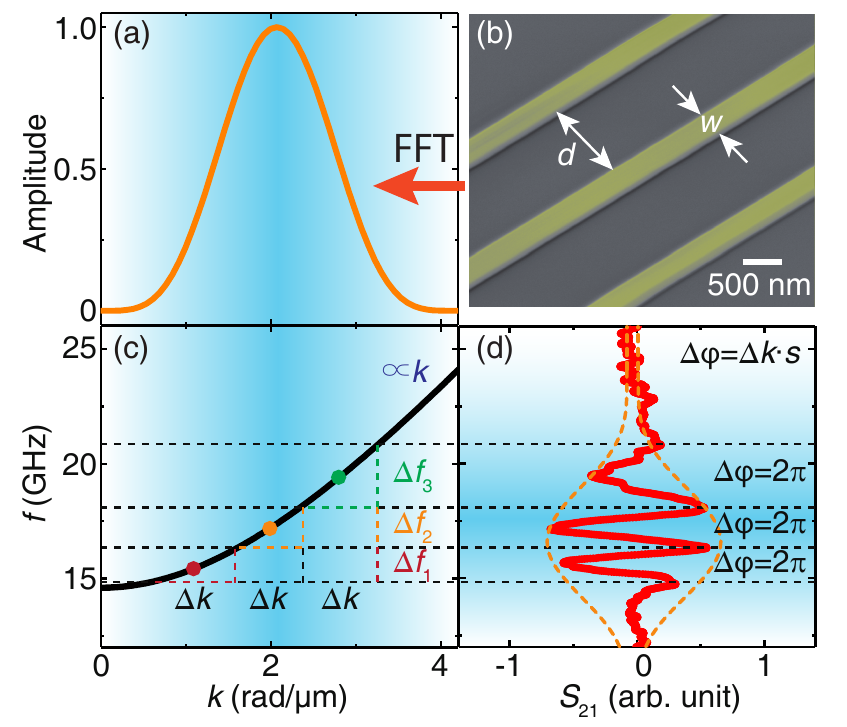}
	\caption{(a) CPW-excited magnetic field wavevector distribution calculated by fast Fourier transformation (FFT) using the dimensions measured in the SEM image of (b) as $w=380$~nm and $d=$1120~nm. (c) AFM spin-wave dispersion plotted for 51~mT, where the background intensity represents the wavevector distribution imposed by CPW as shown in (a). Three equivalent wavevector segments $\Delta k$ project into different frequency spans $\Delta f_{1}$, $\Delta f_{2}$ and $\Delta f_{3}$ in accordance with the dispersion. The frequency spans manifests themselves as peak-to-peak separation in the measured transmission spectrum $S_{21}$ in (d). These frequency spans correspond to a phase change of $2\pi$ after propagation over a certain distance $s$. The orange dashed curve in (d) defines effective excitation envelope corresponding to the wavevector distribution of CPW in (a). Horizontal and vertical black dashed lines are guide for the eyes.}
	\label{fig3}
\end{figure}

\begin{figure}[hb]
	\includegraphics[width=86mm]{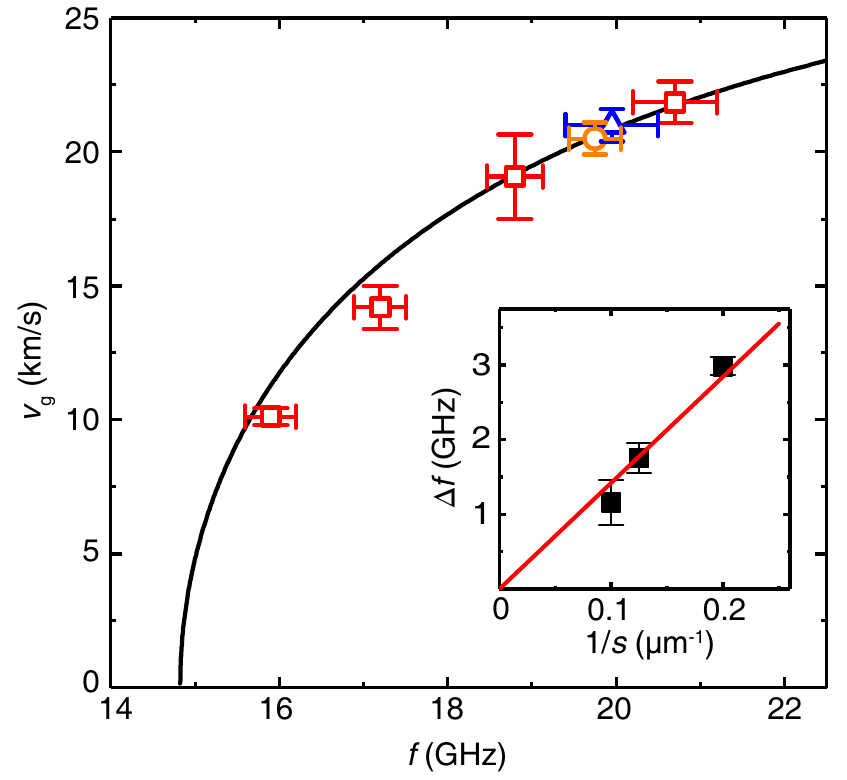}
	\caption{(a) AFM spin-wave group velocities for different frequencies extracted from the transmission spectra $S_{\rm 21}$ at 51~mT. The values are given by the slope from a linear fit of $\Delta f$ versus $1/s$ based on Eq.~\ref{vg1}. The inset shows an example for the linear fitting of the measurement data around 17.2~GHz for three different propagation distances $s=5~\mu$m, $8~\mu$m and $10~\mu$m. Red open squares are data obtained from the samples with the CPW design shown in Fig.~\ref{fig3}(b). Orange circles and blue open triangles are data obtained from the samples with two other CPW designs with slightly different dimensions as described in Sec.~IV in the SM~\cite{SI}. The solid line is the calculation taking the exchange length by fitting of our data $a_{\rm ex}$=1.7~$\rm \AA$} 
	\label{fig4}
\end{figure}

In the following, we present experimental demonstration of high-velocity propagation of spin waves in $\alpha$-Fe$_2$O$_3$ with all-electrical excitation and detection. The DMI-induced small canted moment can couple with an external microwave field and therefore provides us an excellent opportunity to excite antiferromagnetic spin waves with conventional coplanar waveguide antennas~\cite{Yu2014,Vla2008,Neu2010,Liu2018,Che2020} fabricated on hematite using e-beam lithography and evaporation. Figures~\ref{fig2}(a-d) show the scanning electron microscope (SEM) images of the measured device with a spin-wave propagation distance $s\,$=$\,8\,\mu$m. Conventional coplanar waveguides with a ground-signal-ground (GSG) design are patterned on a 0.5-mm-thick $\alpha$-Fe$_2$O$_3$ crystal with e-beam lithography and connected to a vector network analyzer (VNA) to excite and detect spin waves. Transmission spectra $S_{21}$ (excitation at CPW1 and detection at CPW2) are measured by the VNA as a function of magnetic field, sweeping from negative to positive values and shown in Figs.~\ref{fig2}(e-g) for different propagation distances ($s$) of $5\,\mu$m, $8\,\mu$m and $10\,\mu$m. Single spectra extracted at 45~mT are plotted together for three different propagation distances in Fig.~\ref{fig2}(h). The transmission signal amplitude decays due to spin-wave damping. With increasing $s$, the observed signal oscillation becomes denser and the number of peaks (marked by orange arrows) increases, for the following reason. The VNA is sensitive to the phase delay between both antennas and the interval between two neighboring peaks corresponds to a phase difference $\Delta\phi=2\pi$. Over a certain propagation distance $s$, the phase change is given by $\Delta\phi=\Delta k\cdot s$, where $\Delta k$ represents a broad wavevector excitation generated by the nanoscale CPW antennas, as shown in Fig.~\ref{fig3}(a) and (b). When $\Delta\phi=\Delta k\cdot s$ reaches $2\pi$, the second peak appears and the corresponding frequency interval $\Delta f$ is observed [Fig.~\ref{fig3}(d)]. Therefore, the spin-wave group velocity $v_{\text{g}}$ can be estimated~\cite{Yu2014,Vla2008,Neu2010} using
\begin{equation}\label{vg1}
	v_{\rm g}=\frac{\partial\omega}{\partial k}\approx \frac{2\pi\Delta f}{\Delta k}=\frac{2\pi\Delta f}{2\pi/s}=\Delta f\cdot s.
\end{equation}
At different parts of the spin-wave dispersion in Fig.~\ref{fig3}(c), the frequency intervals $\Delta f$ differ ($\Delta f_{1}\neq\Delta f_{2}\neq\Delta f_{3}$) in spite of an identical wavevector increment $\Delta k=2\pi/s$. The higher-frequency part of the dispersion presents a steeper slope, $\Delta f_{3}>\Delta f_{2}>\Delta f_{1}$ as observed in Fig.~\ref{fig3}(d). This indicates an increasing group velocity according to Eq.~\ref{vg1} given a fixed propagation distance $s$. Apart from the phase oscillation, the amplitude envelope (orange dashed arrows) is determined by the broad $k$ excitation (blue shadow region) of the nanoscale CPW as characterized in Fig.~\ref{fig3}(a). The frequency interval $\Delta f_2$ located at the center of the envelope around 17.5~GHz therefore exhibits the largest oscillation amplitude. Based on Eq.~\ref{vg1}, we can extract the average group velocity at about 17.5~GHz (orange dot in Fig.~\ref{fig3}(c)) from a linear fitting of the measured frequency interval $\Delta f$ as a function of $1/s$, as shown in the inset of Fig.~\ref{fig4}. The slope of the fitted red line yields a group velocity of about 14.2~km/s. Following this method, we extract group velocities at different frequency bands and plot them in Fig.~\ref{fig4} as the red open squares. The data acquisition from linear fittings of $\Delta f$ versus $1/s$ for frequencies of 15.9~GHz, 18.8~GHz and 20.7~GHz are presented in the Sec.~IV and V in SM~\cite{SI}. Two additional devices with slightly modified CPWs (Type 2 and Type 3) were also measured (see Sec.~IV and V in SM~\cite{SI}) and group velocities obtained from these two samples are plotted as the orange circle and blue open triangle in Fig.~\ref{fig4}. From the distance-dependent measurements, we extract a decay length of about 10 $\mu$m for coherent AFM spin waves (see Sec.~VI in SM~\cite{SI}). 

The group velocities extracted from different samples plotted in Fig.~\ref{fig4} increase asymptotically toward a saturated group velocity (linear range in dispersion) of $\gamma\mu_0H_{\text{ex}}a_{\rm ex}$. The group velocity as a function of frequency can be derived from the dispersion in Eq.~\ref{AFM_dispersion}. By fitting our data, we obtain an exchange stiffness length $a_{\rm ex}=1.7~\rm\AA$, which could also be considered as an effective lattice constant in the 1D spin-chain model. This value corresponds to an AFM exchange stiffness~\cite{Stohr2004} of about 10~pJ/m and a saturated velocity~\cite{Samu1970} of 30.2~km/s (see Sec.~II in SM~\cite{SI}). To approach the saturated velocity, we fabricate even smaller CPW antennas with larger wavevector $k$ ($\sim$5.2~rad/$\mu$m, see Sec.~VII in SM~\cite{SI}). However, we do not observe clear signal oscillations in transmission spectra, which we attribute to the impedance mismatch as further down-scaling the microwave antennas~\cite{Ciu2016,Luca2019}. For micrometer-scale CPW antennas with small wavevectors ($\sim$1.0 rad/$\mu$m, see Sec.~VII in SM~\cite{SI}), we again do not observe oscillating transmission signals owing to a low group velocity at the low-$k$ limit.

In summary, we experimentally observed the coherent propagating AFM spin waves at room temperature in a single-crystal $\alpha$-Fe$_2$O$_{\rm 3}$ film. Over a long distance of 10 $\mu$m, the coherence of AFM spin waves can still be detected with a high group velocity of up to 22.5~km/s. With measurements using CPW antennas of different propagation distances, the AFM spin-wave dispersion could be indirectly characterized via the relationship between group velocities and frequencies. These data could be accounted for with a theoretical model that takes into account exchange, DM, Zeeman and anisotropy energy. The AFM exchange stiffness length is estimated to be about 1.7 $\rm \AA$. Additional measurements under different configurations between the canted moments and wavevector directions present almost the same frequencies and group velocities, verifying the degenerate AFM spin-wave dispersion shown in Fig.~\ref{fig1} (see Sec.~VIII in SM~\cite{SI}). High-velocity coherent propagating AFM spin waves is suggestive of great prospects for coherent AFM magnonics.\\

\begin{acknowledgments}

The authors thank M. Elyasi, P. Gambardella, K. Yamamoto, and S. Maekawa for their helpful discussions. We wish to acknowledge the support by the National Key Research and Development Program of China Grants No. 2022YFA1402801, NSF China under Grant Nos. 12074026, 52225106, and U1801661, China Scholarship Council (CSC) under Grant No. 202206020091, and Shenzhen Institute for Quantum Science and Engineering, Southern University of Science and Technology (Grant No. SIQSE202007).

\end{acknowledgments}
	
\bibliographystyle{unsrt}

\end{document}